	      \documentstyle[epsfig,amsmath,amssymb]{mn}

\newif\ifAMStwofonts

\ifoldfss
  \ifCUPmtlplainloaded \else
    \NewTextAlphabet{textbfit} {cmbxti10} {}
    \NewTextAlphabet{textbfss} {cmssbx10} {}
    \NewMathAlphabet{mathbfit} {cmbxti10} {} 
    \NewMathAlphabet{mathbfss} {cmssbx10} {} 
  \fi
  \ifAMStwofonts
    \ifCUPmtlplainloaded \else
      \NewSymbolFont{upmath} {eurm10}
      \NewSymbolFont{AMSa} {msam10}
      \NewMathSymbol{\upi}     {0}{upmath}{19}
      \NewMathSymbol{\umu}     {0}{upmath}{16}
      \NewMathSymbol{\upartial}{0}{upmath}{40}
      \NewMathSymbol{\leqslant}{3}{AMSa}{36}
      \NewMathSymbol{\geqslant}{3}{AMSa}{3E}

       \let\le=\leqslant
      \let\geq=\geqslant \let\ge=\geqslant
    \fi
  \fi
\fi 

\ifnfssone
  \newmathalphabet{\mathit}
  \addtoversion{normal}{\mathit}{cmr}{m}{it}
  \addtoversion{bold}{\mathit}{cmr}{bx}{it}
  \newmathalphabet{\mathbfit} 
  \addtoversion{normal}{\mathbfit}{cmr}{bx}{it}
  \addtoversion{bold}{\mathbfit}{cmr}{bx}{it}
  \newmathalphabet{\mathbfss} 
  \addtoversion{normal}{\mathbfss}{cmss}{bx}{n}
  \addtoversion{bold}{\mathbfss}{cmss}{bx}{n}
  \ifAMStwofonts
    \ifCUPmtlplainloaded \else
      \UseAMStwoboldmath
      \makeatletter
      \new@mathgroup\upmath@group
      \define@mathgroup\mv@normal\upmath@group{eur}{m}{n}
      \define@mathgroup\mv@bold\upmath@group{eur}{b}{n}
      \edef\UPM{\hexnumber\upmath@group}
      \new@mathgroup\amsa@group
      \define@mathgroup\mv@normal\amsa@group{msa}{m}{n}
      \define@mathgroup\mv@bold\amsa@group{msa}{m}{n}
      \edef\AMSa{\hexnumber\amsa@group}
      \makeatother
      \mathchardef\upi="0\UPM19
      \mathchardef\umu="0\UPM16
      \mathchardef\upartial="0\UPM40
      \mathchardef\leqslant="3\AMSa36
      \mathchardef\geqslant="3\AMSa3E

       \let\le=\leqslant
      \let\geq=\geqslant \let\ge=\geqslant
    \fi
  \fi
\fi 

\ifnfsstwo
  \DeclareMathAlphabet{\mathbfit}{OT1}{cmr}{bx}{it}
  \SetMathAlphabet\mathbfit{bold}{OT1}{cmr}{bx}{it}
  \DeclareMathAlphabet{\mathbfss}{OT1}{cmss}{bx}{n}
  \SetMathAlphabet\mathbfss{bold}{OT1}{cmss}{bx}{n}
  \ifAMStwofonts
    \ifCUPmtlplainloaded \else
      \DeclareSymbolFont{UPM}{U}{eur}{m}{n}
      \SetSymbolFont{UPM}{bold}{U}{eur}{b}{n}
      \DeclareSymbolFont{AMSa}{U}{msa}{m}{n}
      \DeclareMathSymbol{\upi}{0}{UPM}{"19}
      \DeclareMathSymbol{\umu}{0}{UPM}{"16}
      \DeclareMathSymbol{\upartial}{0}{UPM}{"40}
      \DeclareMathSymbol{\leqslant}{3}{AMSa}{"36}
      \DeclareMathSymbol{\geqslant}{3}{AMSa}{"3E}

       \let\le=\leqslant
      \let\geq=\geqslant \let\ge=\geqslant
    \fi
  \fi
\fi 

\ifCUPmtlplainloaded \else
  \ifAMStwofonts \else 
    \def\upi{\pi}
    \def\umu{\mu}
    \def\upartial{\partial}
  \fi
\fi

\title[WMAP-NVSS: detection of ISW and DE constraints]{Cross-correlation of the CMB and radio galaxies
in real, harmonic and wavelet spaces: detection of the integrated Sachs-Wolfe effect and dark energy constraints}

\author[Vielva et al.]
	{P.~Vielva$^{1}$
	E.~Mart{\'\i}nez-Gonz{\'a}lez$^{2}$ and M.~Tucci$^{2, 3}$\\
	$^{1}$Physique Corpusculaire et Cosmologie, Coll\`ege de France,
        11 pl. M. Berthelot, F-75231, Paris Cedex 5, France \\
	$^{2}$Instituto de F{\'\i}sica de Cantabria (CSIC - UC),
	Avda. Los Castros s/n, 39005, Santander, Spain \\
	$^{3}$Astrophysics Group, The Blackett Laboratory, Imperial College, London SW7 2AZ UK\\
	\hspace{0.2cm}e-mails : vielva@cdf.in2p3.fr, martinez@ifca.unican.es, m.tucci@imperial.ac.uk
        }
\date{\today}

\begin{document}

\maketitle

\label{firstpage}

\begin{abstract}
We report the first detection of the Integrated Sachs-Wolfe effect (ISW) 
in wavelet space, at scales in the sky  around $\theta \approx 7^\circ$ 
with a significance $\approx 3.3\sigma$, by cross-correlating the Wilkinson 
Microwave Anisotropy Probe (WMAP) first-year data and the NRAO VLA Sky 
Survey (NVSS).
In addition, we present a detailed comparison among the capabilities of 
three different techniques for two different objectives: to detect the 
ISW and to put constraints in the nature of the dark energy. The three 
studied techniques are: the cross-angular power spectrum (CAPS, \emph{harmonic space}), 
the correlation function (CCF, \emph{real space}) and the covariance 
of the Spherical Mexican Hat Wavelet (SMHW) coefficients (CSMHW, \emph{wavelet space}).
We prove that the CSMHW is expected to provide a higher detection (in terms of 
the signal-to-noise ratio) of the ISW effect for a certain scale. On the other hand,
the detection achieved by the CAPS is the lowest one, being the signal-to-noise 
ratio dispersed among a wide multipole range. The CCF provides an 
intermediate detection level. This prediction has been corroborated by the analysis 
of the data. The SMHW analysis shows that the cross-correlation signal is caused 
neither by systematic effects nor foreground contamination.
However, by taking into account the information encoded in all the multipoles/scales/angles,
the CAPS provides slightly better constraints than the SMHW in the cosmological
parameters that define the nature of the dark energy. The limits provided by the CCF are wider
than for the other two methods, although the three of them give similar confidence levels.
Two different cases have been studied: 1) a flat $\Lambda$CDM universe and 2) a flat universe
with an equation of state parameter that, although does not change with time, could take values
different from -1.
In the first case, the CAPS provides (for a  bias value of  $b = 1.6$) 
$\Omega_{\Lambda} = 0.73^{+0.11}_{-0.14}$ (at $1\sigma$ CL). Moreover, the CAPS 
rejects the range $\Omega_{\Lambda} < 0.1$ at $\approx 3.5\sigma$, which is the highest 
detection of the dark energy reported up to date.
In the second case, the CAPS gives $\Omega_{DE} = 0.70^{+0.12}_{-0.20}$ and 
$w = -0.75^{+0.32}_{-0.41}$ (at $1\sigma$ CL).
This is the first estimation of the equation of state of the dark energy made through the 
cross-correlation of the CMB and the nearby galaxy density distribution.  
It also provides an independent estimation from the one made by the WMAP team using CMB and LSS.

\end{abstract}


\section{Introduction}
During the last decade, the study of the anisotropies of the cosmic 
microwave background (CMB) is being (probably) the most important 
source for understanding the universe. 
Since the first all-sky detection reported by COBE-DMR in 1992 
(Smoot et al. 1992), many ground and balloon based experiments have 
been measuring the CMB anisotropies, drawing (together with other 
astrophysical data like those provided by supernova) a flat universe 
whose evolution is dominated by the dark energy. This general picture 
has been recently confirmed by the first-year results of the NASA WMAP 
satellite (Bennett et al. 2003a, Spergel et al 2003), providing a strong 
confirmation of the fiducial $\Lambda$CDM model. 

At this point, one of the most interesting questions is the confirmation 
of this fiducial model by other independent analysis. In that sense, the 
Integrated Sachs-Wolfe effect (ISW, Sachs \& Wolfe, 1967) --that is 
produced by the time variation of the gravitational potential, in the 
linear regime-- plays a crucial role since it provides either a direct 
indication of the presence of dark energy in the case of a flat universe 
or else the existence of spatial curvature (Peebles \& Ratra 2003).
Since WMAP has established strong constraints in the flatness of the 
universe (Bennett et al. 2003a), the detection of the ISW 
will imply the detection of the dark energy. 

Many different cosmological models can be included within the dark energy 
concept: the standard inflationary model dominated by a cosmological 
constant (with an equation of state parameter $w = -1$; $p \equiv w\rho$) 
and alternative models dominated by a dark energy whose energy density 
is spatially inhomogeneous with negative pressure and evolving with time, 
like topological defects, quintessence models and phantom models (see for 
instance Melchiorri 2004 and references therein).  
However, there is not yet a convincing explanation for the origin and
nature of the dark energy within the practical physics framework.
Obviously, any information extracted from cosmological data, that
could help to characterise it, is very valuable (for phenomenological 
aspects of dark energy see for instance Corasaniti 2004).

Crittenden \& Turok (1996) firstly suggested that the ISW could be detected by 
cross-correlating the CMB with the nearby galaxy density distribution. 
The first attempt to detect the cross-correlation of the CMB and the galaxy 
density distribution was done by Boughn \& Crittenden (2002) using the COBE-DMR 
map and the NRAO VLA Sky Survey (NVSS, Condon et al. 1998) data. In that work, 
no cross-correlation was found, concluding that a future experiment with better 
sensitivity and resolution was required in order to detect it.
The first-year WMAP data has provided a unique tool to perform such correlation. 
Several groups have reported ISW detection at different significant levels. 
Boughn \& Crittenden (2004) have correlated the WMAP data with two different 
tracers of the nearby universe: the hard X-ray data provided by the HEAO-1 satellite 
(Boldt 1987) and the NVSS data. They find a statistical significance of 1.8 - 2.8 
$\sigma$ at scales lower than $3^\circ$.
Nolta et al. (2004) have done an independent analysis of the WMAP and NVSS 
cross-correlation obtaining $\Omega_{\Lambda} > 0$ at 95\%. Fosalba \& Gazta\~naga 
(2003) find a cross-correlation between WMAP and the APM Galaxy survey (Maddox et 
al. 1990) at 98.8\% significance level at scales $\approx 4^\circ$ - $10^\circ$. 
Fosalba et al. (2003) cross-correlated WMAP with the 
Sloan Digital Sky Survey (SDSS, Abazajian et al. 2004) finding a positive ISW 
signal at the $3\sigma$ level and determining $\Omega_\Lambda = 0.69 - 0.86$ 
($2\sigma$ CL). These two data sets were also analysed by Scranton et al. 
(2003)\footnote{See Afshordi et al. (2004) for a critical discussion of the 
analysis done in Fosalba \& Gazta\~naga (2003), Fosalba et al. (2003) and Scranton 
et al. (2003)}. Afshordi et al. (2004) have cross-correlated the galaxies from the 
Near-IR Two Micron All Sky Survey (2MASS, Skrutskie et al. 1997) finding an ISW 
signal at 2.5$\sigma$ level.

All the previous works have performed the cross-correlation of the WMAP data and 
the nearby universe tracers using the cross-angular power spectrum (CAPS, e.g. 
Afshordi et al. 2004) and the correlation function (CCF, e.g. Boughn \& Crittenden, 
2002, 2004, Nolta et al. 2004).
In the present paper, we propose an alternative to the \emph{harmonic} and \emph{real} \emph{spaces}
to perform the correlation of the WMAP and NVSS data sets: the \emph{wavelet space}. 
The motivation for using wavelets is clear since the ISW signal provided by the 
CMB-galaxies cross-correlation takes place at scales between $2 ^\circ- 10^\circ$ 
(Afshordi 2004). As it is well known, wavelets are very suitable tools for 
detecting signals with a characteristic scale: by filtering the data at a given 
scale, those structures having that scale are amplified, allowing for a most optimal 
detection. Therefore, the wavelet analysis of the cross-correlation will concentrate
almost all the signal of the ISW detection in a small scale range, allowing for 
a fast and direct measure of its signal-to-noise. This is not possible neither for 
the CAPS nor the CCF, where all the information spread out among all the multipoles/angles 
must be considered to estimate the total signal-to-noise. This effect is even more
important for the CAPS than for the CCF.

Since we expect that the common structures 
in CMB and galaxy surveys have roundish rather than elongated shapes, we propose 
the Spherical Mexican Hat Wavelet (SMHW) to study the cross-correlation.
A natural statistic of the cross-correlation in wavelet space is the covariance of the
SMHW coefficients (CSMHW) at each scale.
Wavelets have been extensively used in many astrophysical and cosmological analyses. 
In the CMB field, they have been used to study the non-Gaussianity of the CMB (Pando 
et al. 1998, Hobson et al. 1999, Aghanim \& Forni, 1999, Tenorio et al. 1999, Barreiro 
et al. 2000, Mukherjee et al., 2000, Barreiro \& Hobson 2001, Cay\'on et al. 2001, 2003, 
Mat{\'\i}nez-Gonz\'alez et al. 2002, Vielva et al. 2004, Starck et al. 2004, Cruz et al. 2005, 
McEwen et al. 2005), the component separation problem (Cay\'on et 
al. 2000, Vielva et al. 2001a, b, 2003) and denoising (Sanz et al. 1999a, b). This work 
is the first application to the cross-correlation of CMB and other data sets. 

In addition to the pure ISW detection, the CAPS, the CCF and the CSMHW can be exploited, 
by comparing the ISW signal with the one provided by different 
cosmological models, to put constraints in the nature of the dark energy: 
not only in the quantity of the dark energy ($\Omega_{DE}$), but also in the equation 
of state parameter ($w$).

The paper is organized as follows. In Section~\ref{sec.data} we describe the analysed 
data sets (WMAP and NVSS). In Section~\ref{sec.tool} the three 
cross-correlation estimators (CAPS, CCF and CSMHW) are presented. 
The expected signal-to-noise ratio for the ISW detection for the CAPS, the CCF and the CSMHW 
is calculated in Section~\ref{sec.motivation}.
The results are given in Section ~\ref{sec.reso} and, finally, in Section ~\ref{sec.conclusion} 
are the conclusions.

\section{The WMAP and NVSS data sets}
\label{sec.data}
The two data sets that have been used in order to perform the CMB-nearby universe 
cross-correlation are the Wilkinson Microwave Anisotropy Probe (WMAP, Bennett et al. 2003a and 
references therein) first-year data and the NRAO VLA Sky Survey (NVSS, Condon et al. 1998).

\begin{figure*}
	\begin{center}
	\includegraphics[angle=90,width=8cm]{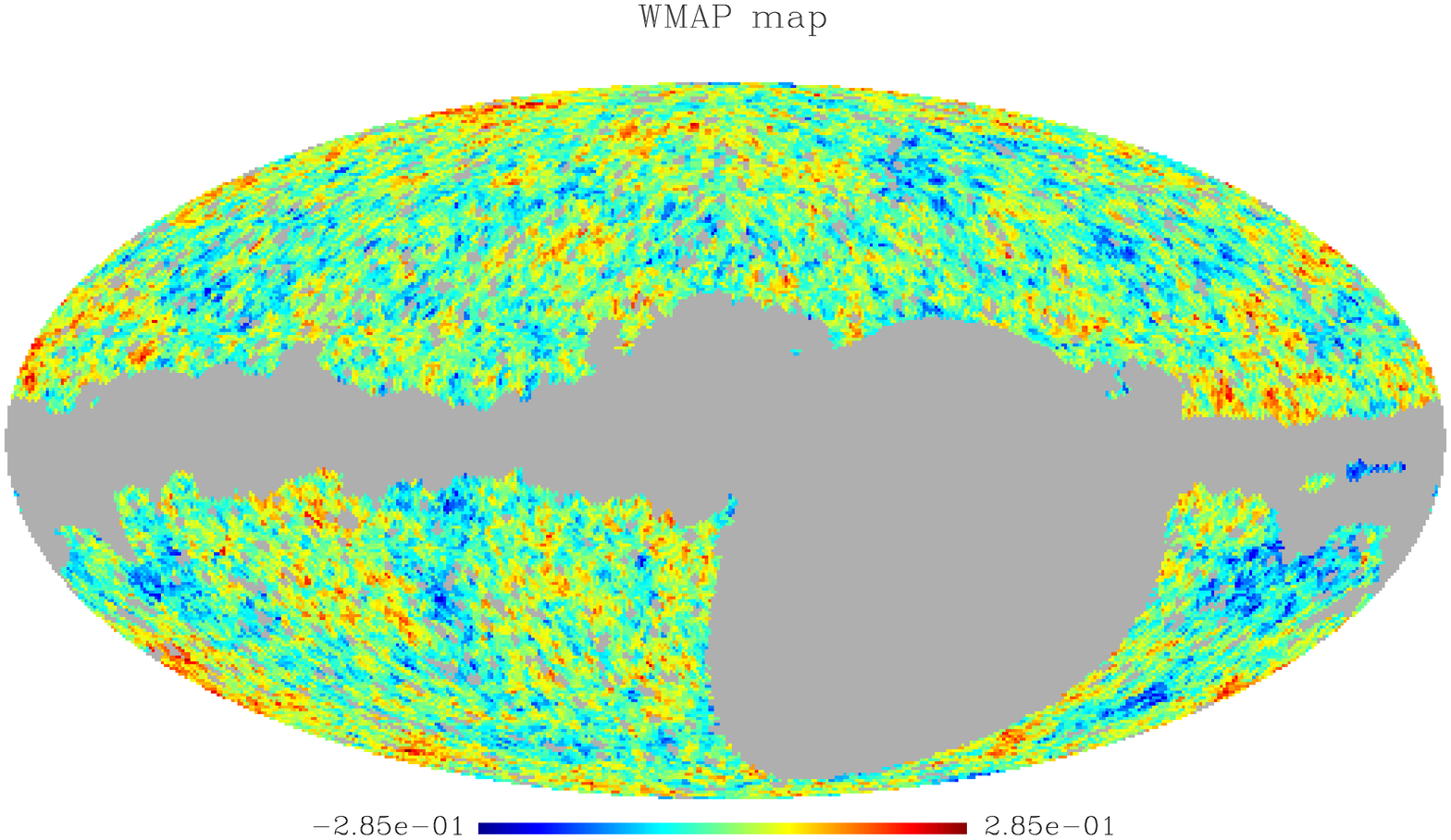}
	\includegraphics[angle=90,width=8cm]{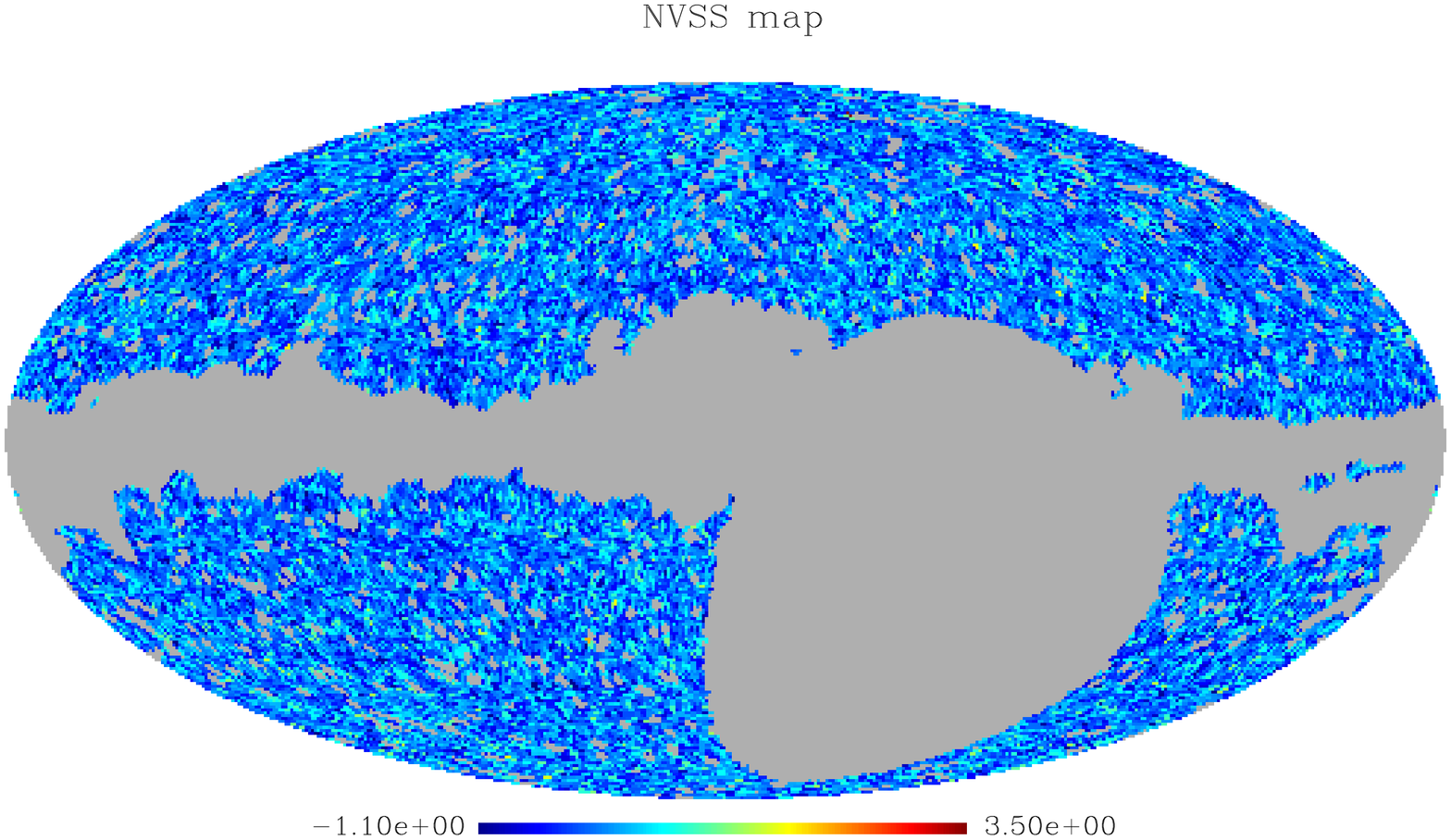}
	\caption{\label{fig.data}Analysed WMAP and NVSS data after the
	application of the \emph{joint mask} and the subtraction of
	the residual monopole and dipole. The maps are represented in 
	the HEALPix scheme, with a resolution parameter  $N_{side} = 64$
	pixel size $\approx$ 55 arcmin).}  
	\end{center}
\end{figure*}

\subsection{WMAP data}

The WMAP radiometers observe at five frequencies: 22.8, 33.0, 40.7, 60.8 and 93.5 GHz, 
having 1, 1, 2, 2 and 4 receivers per frequency band, respectively. All the papers, data 
and products generated by the WMAP team can be found in the Legacy Archive for Microwave 
Background Data Analysis (LAMBDA) Web site\footnote{Available at http://cmbdata.gsfc.nasa.gov}. 
The WMAP maps are presented in the HEALPix scheme (G\'orski et al. 2005) at the 
$N_{side} = 512$ resolution parameter.
The WMAP team and other groups have proposed different CMB maps obtained from the WMAP 
data. In this work, we have used the map proposed by the WMAP team and already used by 
other groups (Komatsu et al. 2003, Vielva et al. 2004, Eriksen et al. 2004, 
Mukherjee \& Wang 2003, Hansen et al. 2004, Cruz et al. 2005) for the study of the non-Gaussianity and the 
isotropy of the CMB. This map is generated (see Bennett et al. 2003b for details and 
Vielva et al. 2004 for a summarized description) as the noise weighted combination of 
all the maps produced by the receivers in which the CMB is the dominant signal (40.7, 60.8 
and 93.5 GHz), after subtraction of the foreground emission and application of the so-called 
``Kp0'' Galactic mask (defined by the WMAP team and where the brightest point sources are 
also masked). Whereas for the non-Gaussianity studies the resolution $N_{side} = 256$ was 
commonly chosen, in the present work we have degraded the combined, corrected and masked map 
down to $N_{side} = 64$ (pixel size $\approx 55$ arcmin). 
The reason is that, as it was pointed out by Afshordi (2004), almost all the signal of the 
ISW is expected to be generated by structures with a scale larger than $2^\circ$. Hence, a 
WMAP resolution of around $1^\circ$ is enough.

\subsection{NVSS data}

The NVSS catalogue covers around 80\% of the sky and has flux and polarization measurements 
for almost 2 million of point sources with a minimum flux $\approx 2.5$ mJy at 1.4 GHz. 
This catalogue has been already used for performing the correlation with the WMAP data 
(Boughn \& Crittenden 2002, 2004 and Nolta et al. 2004). 
In this work, we have represented the point source catalogue in the HEALPix scheme, also at 
the $N_{side} = 64$ resolution.
Only the sources above 2.5 mJy have been used, which represents the 50\% completeness 
(Condon et al. 1998). As it was pointed out by Boughn \& Crittenden (2002) and Nolta et al. 
(2004), the mean density of point sources varies as a function of the declination. This 
systematic effect was corrected, by imposing that the mean of the galaxy density at each 
iso-latitude band is the same. The iso-latitude bands are defined taking into account the 
change in the RMS noise levels in the NVSS survey (see Figure 10 of Condon et al. 1998).
Alternative strategies (as the one proposed by Nolta et al. 2004) were also considered, proving that
results are not sensitive to the particular correction procedure.

As it was said before, the NVSS catalogue covers around 80\% of the sky: for equatorial 
declination lower than $-50^\circ$ there are not observations and, within the 
range $-37^\circ > \delta > -50^\circ$ the coverage is not good enough. Hence, we only consider 
sources with an equatorial declination $\delta \geq -37^\circ$. With all these constraints, 
we have a galaxy distribution map of $\approx 1600000$ radio sources with an average number 
of 40.4 counts per pixel.

In Figure~\ref{fig.data} we have plotted the two maps to be analysed: WMAP (left) and NVSS (right). 
Both are in Galactic coordinates and the \emph{joint mask} (Kp0 + $\delta < -37^\circ$) has been 
applied. The residual monopole and dipole outside the mask have been removed.

\subsection{Simulations}

We have also performed realistic simulations in order to carry out the analysis. 1000 Gaussian 
simulations of the WMAP data have been done, following the concordance cosmological model given 
by the Table 1 of Spergel et al. 2003 ($\Omega_\Lambda = 0.71$, $\Omega_m = 0.29$, $\Omega_b = 
0.047$, $H_0 = 72$, $\tau = 0.166$, $n = 0.99$) and using the CMBFAST code (Seljak \& Zaldarriaga 1996).
For each realization, we have simulated all the WMAP data measured by the receivers at 
40.7, 60.8 and 93.5 GHz, they have been convolved with the real beams provided in the LAMBDA 
Web site, the anisotropic WMAP noise was added, the maps were combined using a noise-weighted average, 
the combined map was degraded to the $N_{side} = 64$ resolution and the \emph{joint mask} was applied. 
We have cross-correlated the 1000 CMB simulations with the NVSS data, in order to evaluate the 
significance level of the cross-correlation obtained from WMAP and NVSS. This is enough to quantify
the covariance matrix associated to random cross-correlations and we have checked that
it is almost independent of the cosmological model used to simulate the CMB.

\section{The cross-correlation estimators: CAPS, CCF and CSMHW}
\label{sec.tool}
As it is well known (e.g. Peebles \& Ratra 2003) for a flat universe where 
the dynamics are dominated by the dark energy, we expect a positive correlation 
between the CMB and the galaxy distribution of the nearby universe ($z \lesssim 1$, 
see for instance Afshordi 2004). 

In this paper we apply three different techniques to study such correlation and
to compare the performance of each of them for the detection of the ISW.
The three studied techniques are the cross-angular power spectrum (CAPS), the 
correlation function (CCF) and the covariance of the SMHW coefficients (CSMHW), 
covering the \emph{harmonic}, \emph{real} and \emph{wavelet} \emph{spaces}.
The CAPS has been already used, for instance, in Afshordi et al. (2004) to estimate
the Sunyaev-Zeldovich, the point sources and the ISW signals 
by cross-correlating the 2MASS infrared source catalogue and WMAP.
The CCF has been more extensively used, for instance in Boughn \& Crittenden (2004) and 
Nolta et al. (2004). We proposed a new technique based in \emph{wavelet space}, the CSMHW.

\subsection{The cross-angular power spectrum}

The observed CAPS of WMAP and NVSS is given by:
\begin{equation}
\label{eq.obcls1}
{\tilde{C}_\ell}{_{\tiny{WN}}} = \left< t_{\ell m} n_{\ell m}^* \right>, 
\end{equation}
where $t_{\ell m}$ and $n_{\ell m}$ are the harmonic coefficients of the WMAP and the NVSS maps
(respectively) and $^*$ denotes complex conjugation.

This CAPS can be compared with the \emph{theoretical} one expected from a given cosmological model:
\begin{equation}
\label{eq.obcls2}
{{\tilde{C}_\ell}^{theo}}{_{\tiny{WN}}} = {p_\ell}^2 ~ {b_W}_\ell ~ {b_N}_\ell ~ M_{\ell, \ell'} ~ {{C_{\ell'}}^{theo}}_{_{\tiny{WN}}},
\end{equation}
where $p_\ell$ is the pixel window function for the corresponding pixelization, 
${b_W}_\ell$ and ${b_N}_\ell$ are the beam window function for WMAP and 
NVSS\footnote{For the NVSS galaxy density map there is no beaming, therefore 
${b_N}_\ell = 1 ~ \forall ~ \ell$}, respectively, $M_{\ell, \ell'}$ is a coupling matrix 
that accounts for the \emph{joint mask} (see Hivon et al. 2002) 
and ${{C_\ell}^{theo}}{_{\tiny{WN}}}$ is the pure CAPS of WMAP and NVSS.
Since $N_{side}=64$ resolution maps are analysed, only multipoles from $\ell=2$ to $\ell=191$ are considered.
As it can be seen in, e.g., Nolta et al. (2004), ${{C_\ell}^{theo}}{_{\tiny{WN}}}$ is given by:
\begin{equation}
\label{eq.cross-cl}
{{C_\ell}^{theo}}_{_{\tiny{WN}}} = 12 \pi \Omega_m {H_o}^2 \int \frac{dk}{k^3} {\Delta_\delta}^2 (k) {F_\ell}^W (k) {F_\ell}^N (k),
\end{equation}
where $\Omega_m$ is the matter density, $H_o$ is the Hubble constant, 
${\Delta_\delta}^2 (k) = k^3 P_\delta(k)/2\pi^2$ is the logarithmic matter power 
spectrum (being $P_\delta(k)$ the matter power spectrum) and ${F_\ell}^W (k)$ 
and ${F_\ell}^N (k)$ are \emph{filter functions} for the CMB and the galaxy 
density distribution, respectively, given by:
\begin{eqnarray}
\label{eq.fls}
{F_\ell}^W (k) & = & \int dz \frac{dg}{dz} j_\ell(k\eta(z)) \\
{F_\ell}^N (k) & = & b \int dz \frac{dN}{dz} D(z) j_\ell(k\eta(z))  
\end{eqnarray}
In the case of ${F_\ell}^W (k)$, the integral is defined from $z = 0$ to $z$ at 
recombination epoch. The integration range for the \emph{filter function} ${F_\ell}^N (k)$ 
is defined, in practice, by the source redshift distribution function $\frac{dN}{dz}$. 
The function $D(z)$ is the linear growth factor for the matter distribution (calculated 
from CMBFAST by computing the transfer function for different redshifts), 
$g \equiv (1 + z)D(z)$ is the linear growth suppression factor, $j_\ell(k\eta(z))$ 
is the spherical Bessel function and $\eta(z)$ is the conformal look-back time. 
The bias factor $b$ is assumed to be redshift independent. 

For the evolution of $D(z)$ we consider the standard model dominated by a 
cosmological constant as well as alternative models dominated by a dark energy
whose energy density is spatially inhomogeneous with negative pressure and 
evolving with time, $\rho = \rho_0 (1+z)^{3(1+w)}$, where $w$ is the equation 
of state parameter defined as the ratio of pressure to density.
Whereas the standard inflationary model assumes $w = -1$ and a homogeneous field 
(the cosmological constant), in general dark energy models are characterised by 
values of $w < 0$. For instance, topological defects can be phenomenologically 
represented by an equation of state parameter $-1/3 \ge w \ge -2/3$ (see e.g. 
Friedland et al. 2003); quintessence models (Wetterich 1988, Caldwell et al. 1998) 
imply $-1 < w < 0$ and phantom models have equation of state parameters 
$w < -1$ (these last models, however, violate the null dominant energy condition; 
see Carroll et al. 2003 for a detailed discussion).

Although a convincing explanation for the origin and nature of the dark 
energy within the framework of particle physics is lacking, the models considered 
in the literature produce in general $w$ varying with time. However, for most dark 
energy models, the equation of state changes slowly with time and
a standard approximation assumes that (at least during a given
epoch) $w$ can be considered
as an effective (and constant) equation of state parameter 
(Wang et al. 2000). Therefore, in this paper, we consider $w$ constant as an 
useful approach to extract fundamental properties of the dark energy.

\subsection{The correlation function}

The observed CCF is given by:
\begin{equation}
\label{eq.obscross-cf}
CCF(\theta) = \frac{1}{N}\sum_{\vec{n}, \vec{n}'}{T(\vec{n}) N(\vec{n}')},
\end{equation}
where $T(\vec{n})$ and $N(\vec{n})$ are the WMAP temperature and NVSS galaxy density (respectively) at position $\vec{n}$.
The sum is extended over all the pixels that are allowed by the \emph{joint mask} and inside a disk
described by $(\theta - \Delta \theta/2, \theta + \Delta \theta/2)$ 
where $\vec n \vec{n}' = cos(\theta)$
and $\Delta \theta = 2^{\circ}$.
As in Nolta et al. (2004) the CCF is evaluated at 10 angles $\theta$ = 1, 2, 3, 5, 7, 9, 10, 13, 15, 17
and 19 degrees.

It can also be compared with the \emph{theoretical} one expected from a given cosmological model:
\begin{eqnarray}
\label{eq.cross-cf}
{CCF}^{theo}(\theta) & = & \Big< \delta T(\vec n) \delta N(\vec{n}') \Big> \\
\nonumber & = & \sum_\ell \frac{2\ell + 1}{4\pi}{P_\ell(\vec n  \vec{n}')} ~ {p_\ell}^2 ~ {b_W}_\ell ~ {b_N}_\ell ~ {{{C_\ell}^{theo}}_{_{\tiny{WN}}}},
\end{eqnarray}
where ${P_\ell(\vec n \vec{n}')}$ are the \emph{Legendre} polynomials and 
${{C_\ell}^{theo}}_{_{\tiny{WN}}}$ is given by Equation~\ref{eq.cross-cl}.

\subsection{The covariance of the wavelet coefficients}

We define the CSMHW at a given scale $R$ as:
\begin{equation}
\label{eq.estimator}
{Cov}_{_{{\tiny{WN}}}} (R)= \frac{1}{N_{R}}\sum_{\vec{p}}{\omega}_{_{{\tiny{T}}}}(R, \vec{p}) {\omega}_{_{{\tiny{N}}}}(R, \vec{p}),
\end{equation}
where ${\omega}_{_{{\tiny{T}}}}(R, \vec{p})$ and ${\omega}_{_{{\tiny{N}}}}(R, \vec{p})$ are the 
SMHW coefficients of WMAP and NVSS (respectively) at position $\vec{p}$.
We propose to study the cross-correlation of the WMAP and NVSS Spherical Mexican Hat Wavelet (SMHW) 
coefficient maps at different scales ($R_1 = 50.0$, $R_2 = 75.0$, $R_3 = 100.0$, $R_4 = 150.0$, 
$R_5 = 200.0$, $R_6 = 250.0$, $R_7 = 300.0$, $R_8 = 400.0$, $R_9 = 500.0$, $R_{10} = 600.0$, 
$R_{11} = 750.0$,  $R_{12} = 900.0$,  $R_{13} = 1050.0$ arcmin) in the angular range where the ISW 
signal is expected to be more important (Afshordi 2004) for CMB-galaxy cross-correlations. 

The SMHW coefficients are obtained by convolving the map with the SMHW at a given scale:
\begin{equation}
{\omega}_{_{{\tiny{X}}}}(R, \vec{p}) = \int d\Omega' \, X(\vec{p} + \vec{p'})
	\Psi_S(\theta', R),
\end{equation}
where $\Psi_S(\theta', R)$ is the SMHW:
\begin{equation}
\label{eqSMHW}
   \Psi_S(y,R) = \frac{1}{\sqrt{2\pi}N(R)}{\Big[1+{\big(\frac{y}{2}\big)}^2\Big]}^2
  \Big[2 - {\big(\frac{y}{R}\big)}^2\Big]e^{-{y}^2/2R^2},
\end{equation}
being $N(R)$ a normalization constant:
\begin{equation}
\label{N}
      N(R)\equiv R{\Big(1 + \frac{R^2}{2} + \frac{R^4}{4}\Big)}^{1/2}.
\end{equation}
The distance on the tangent plane is given by $y$ that is related 
to the latitude angle ($\theta$) through:
\begin{equation}
\label{y}
      y\equiv 2\tan \frac{\theta}{2}.
\end{equation}
The SMHW is obtained from the Euclidean Mexican Hat Wavelet (MHW) following the 
stereographic projection suggested by Antoine \& Vanderheynst (1998). The reader 
is referred to the work of Mart{\'\i}nez-Gonz\'alez et al. (2002) where this 
projection is described for the SMHW, as well as its properties.

In Equation~\ref{eq.estimator}, the sum $\sum_{\vec{p}}$ is extended over all the 
pixels that are not masked ($N_{R}$). As it was already discussed in Vielva et al. 
(2004), since we are convolving a map with a large mask (the \emph{joint mask} 
covers 43\% of the sky) with the SMHW at a given scale $R$, we are introducing into 
the cross-correlation estimator ${Cov}_{_{{\tiny{WN}}}} (R)$ a large number of 
pixels --those close to the border of the mask-- highly affected by the zero value 
of the mask, which implies a loss of efficiency. For that reason, at a given scale 
R, we \emph{exclude} from the calculation of ${Cov}_{_{{\tiny{WN}}}} (R)$ all the 
pixels with a strong contamination from the \emph{joint mask}. Whereas in Vielva et 
al. (2004) those pixels that were closer than $2.5R$ to any one of the pixels of the 
\emph{joint mask} where excluded, for the present analysis we exclude only those pixels 
closer than $1.0R$ to the \emph{joint mask}. There are two reasons for that: first, the
 mask used in the present work is much larger than the one used in Vielva et al. (2004) 
and second, the number of pixels is lower, since the HEALPix resolution is 3 times lower.
These two facts make the available number of data much smaller (for all the scales) than 
the ones in Vielva et al. (2004) and, therefore, we must be less restrictive in order 
to construct the 13 \emph{exclusion masks}. However, we have checked that the ISW is also 
detected with the more restrictive \emph{exclusion masks} proposed by Vielva et al. (2004).

The curve defined by Equation~\ref{eq.estimator} can be easily compared with the theoretical 
CSMHW between the CMB and galaxy density distribution: the ${Cov}_{_{{\tiny{WN}}}} (R)$ 
is nothing more than the mean value of a map $M(R) = {\omega}_{_{{\tiny{T}}}}(R) {\omega}_{_{{\tiny{N}}}}(R)$, 
where the reference to the position $\vec{p}$ has been removed for simplicity. 
It is straightforward to show that the theoretical wavelet cross-correlation can be written as:
\begin{equation}
\label{eq.theo}
{{Cov}^{theo}}_{_{{\tiny{WN}}}} (R)= \sum_\ell \frac{2\ell + 1}{4\pi}{C_{\ell_{M}}(R)}
\end{equation}
where $C_{\ell_{M}}(R)$ is the angular power spectrum of the map $M(R)$ and it is given by:
\begin{equation}
\label{eq.cls}
C_{\ell_{M}} (R) = {p_\ell}^2 ~ {s^2}_\ell (R) ~ {b_W}_\ell ~ {b_N}_\ell ~ {{C_\ell}^{theo}}_{_{\tiny{WN}}},
\end{equation}
where, as before,  $p_\ell$ is the pixel window function for the corresponding pixelization, 
${b_W}_\ell$ and ${b_N}_\ell$ are the beam window function for WMAP and NVSS, 
${{C_\ell}^{theo}}_{_{\tiny{WN}}}$ is the pure CAPS of WMAP and NVSS and 
$s_\ell (R)$ are the spherical harmonic coefficients of the SMHW at scale $R$ 
(there is no dependence with $m$, since the SMHW is isotropic).

\noindent Summarising, the comparison of the experimental curves
given by Equations~\ref{eq.obcls1},~\ref{eq.obscross-cf} and~\ref{eq.estimator} 
with the theoretical prediction given by Equation~\ref{eq.obcls2},~\ref{eq.cross-cf} 
and~\ref{eq.theo} can be done, by computing the 
\emph{filter functions} and the pure CAPS for different cosmological models.

\section{Comparison of different techniques: motivation for using wavelets}
\label{sec.motivation}

The motivation for using wavelets for the detection of the ISW signal
through the CMB-galaxies  
cross-correlation is given by the fact that such correlation has his
maximum contribution  
in the scale range between $2^{\circ}$ and $10^{\circ}$ (see Afshordi 2004).
As it is well known, wavelets are ideal tools for detecting signals
(embedded in a given  
background) with a characteristic scale.
The convolutions of the data with a wavelet with a certain scale, amplifies those 
structures with such scale, allowing for a most optimal detection.

\begin{figure}
	\begin{center}
	\includegraphics[width=8cm]{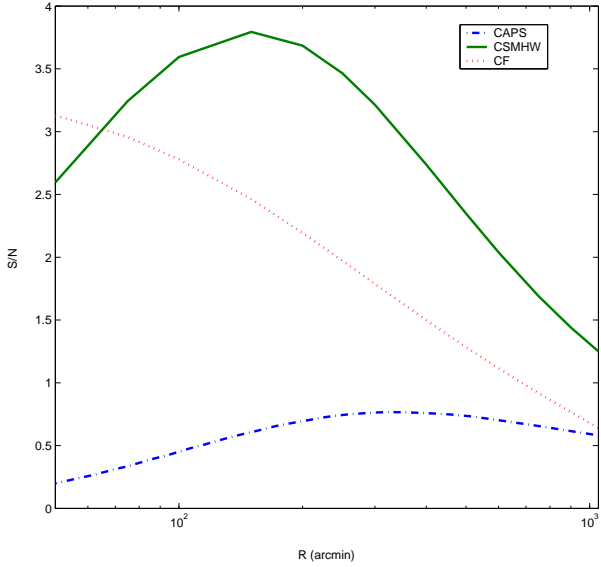}
	\caption{\label{fig.s2n} The expected signal-to-noise ratio of
	  the CMB--radio sources cross-correlation according to the
	  three studied techniques: the CSMHW (solid curve), the
	  CAPS (dotted--dashed curve) and the CCF (dotted
	  curve). The cosmological parameters are chosen according to
	  the concordance model.}

	\end{center}
\end{figure}

In the present Section, we compare the expected
signal--to--noise ratio of the cross-correlated signal
between CMB fluctuations and the radio sources density distribution
as a function of the
angular scale for the estimators described in the previous section
(see Figure~\ref{fig.s2n}): the cross--angular spectrum
${C_\ell}_{_{\tiny{WN}}}$ (CASP, Eq.~\ref{eq.cross-cl}); the
correlation function (CCF, Eq.~\ref{eq.cross-cf}) and the covariance of
wavelet coefficients ${Cov}_{\tiny{WN}}(R)$ (CSMHW, Eq.~\ref{eq.theo}).
This is done for an ideal free-noise experiment with full-sky coverage
and a resolution $N_{side} = 64$ (and assuming no beaming, 
i.e. ${b_W}_\ell = {b_N}_\ell = 1$).

The signal--to--noise level can be estimated as the ratio between the
cross--correlation estimator and its dispersion. As it is
straightforward to show, the dispersion of the CAPS for two Gaussian
random fields is given by:
\begin{equation}
\label{eq.error-cls}
\Delta {{C_\ell}^{theo}}_{_{\tiny{WN}}} = \sqrt{ \frac{ {{C_\ell}^{theo}}_{_{\tiny{WN}}}^2 + {{C_\ell}^{theo}}_{_{\tiny{W}}} {{C_\ell}^{theo}}_{_{\tiny{N}}}}{2\ell + 1} }
\end{equation}
where ${{C_\ell}^{theo}}_{_{\tiny{W}}}$ and
${{C_\ell}^{theo}}_{_{\tiny{N}}}$ are the angular power spectra for
the CMB and the
radio sources density distribution, respectively.

Using Equation~\ref{eq.theo} and~\ref{eq.error-cls}, 
we find that the dispersion of the correlation of the CSMHW is given by:
\begin{eqnarray}
\label{eq.error-wave}
\nonumber \Delta {{Cov}^{theo}}_{_{{\tiny{WN}}}} (R)   & = & \\ \nonumber
\sqrt{ \Big< {{{Cov}^{theo}}_{_{{\tiny{WN}}}} (R)}^2 \Big> 
- \Big< {{Cov}^{theo}}_{_{{\tiny{WN}}}} (R) \Big>^2} & = & \\
\sqrt{ \sum_\ell{  \frac{2\ell + 1}{16\pi^2}  {p_\ell}^4  ~ {s_\ell}^4 (R) 
( {{C_\ell}^{theo}}_{_{\tiny{WN}}}^2 + {{C_\ell}^{theo}}_{_{\tiny{N}}} {{C_\ell}^{theo}}_{_{\tiny{W}}} ) } } & &
\end{eqnarray}
Equivalently, the dispersion of the CCF is given by:

\begin{eqnarray}
\label{eq.error-cf}
\nonumber \Delta  CCF^{theo}(\theta)  & = & \\ \nonumber
\sqrt{ \Big< {CCF^{theo}(\theta)}^2 \Big> 
- \Big< CCF^{theo}(\theta) \Big>^2} & = & \\
\sqrt{ \sum_\ell{  \frac{2\ell + 1}{16\pi^2}  {p_\ell}^4 ~ {P_\ell}^2(\vec n \vec{n}')
( {{C_\ell}^{theo}}_{_{\tiny{WN}}}^2 + {{C_\ell}^{theo}}_{_{\tiny{N}}} {{C_\ell}^{theo}}_{_{\tiny{W}}} ) } } & &
\end{eqnarray}

Figure~\ref{fig.s2n} shows that the CSMHW gives the highest
signal-to-noise ratio ($\approx 3.5$), at scales of $\approx
4^{\circ}$. As expected, the maximum significance for the CCF ($\approx3$) 
is reached at the smallest scale.
The signal-to-noise ratio for the CAPS is significantly lower than for the other
ones and always below 1, with a shape nearly flat ($\approx 0.6$).

Let us recall two important points. First, the values of the signal-to-noise ratio curves for both, the 
CSMHW coefficients and  the CCF have a very high scale-to-scale correlation, 
whereas there is no correlation in the case of the CAPS. 
In principle,
by taking into account all the scales and the full correlations, one should expect to get 
comparable detection levels in the
three cases. However, we remark that, for a certain scale, 
the detection of the ISW given by the CSMHW is proven to be the highest. 
In other words, the whole information of the ISW has been extremely concentrated and amplified
in a very narrow range of angular scales. This is undoubtedly very useful in a practical situation, 
like the one addressed in this paper.

Second, in the case of a noisy experiment, the signal-to-noise ratio should be lower than 
the one obtained under our ideal conditions.
This will be much more important for an estimator as the CCF, since it has its maximum amplification 
at the smallest scale.
On the contrary, the CSMHW and the CAPS are almost unaltered in this case, 
since the maximum amplification is reached at intermediate angular scales, where
the noise is almost negligible. The same argument is also applied if beaming is included.

\section{Results}
\label{sec.reso}
\begin{figure*}
	\begin{center}
	\includegraphics[angle=270,width=16cm]{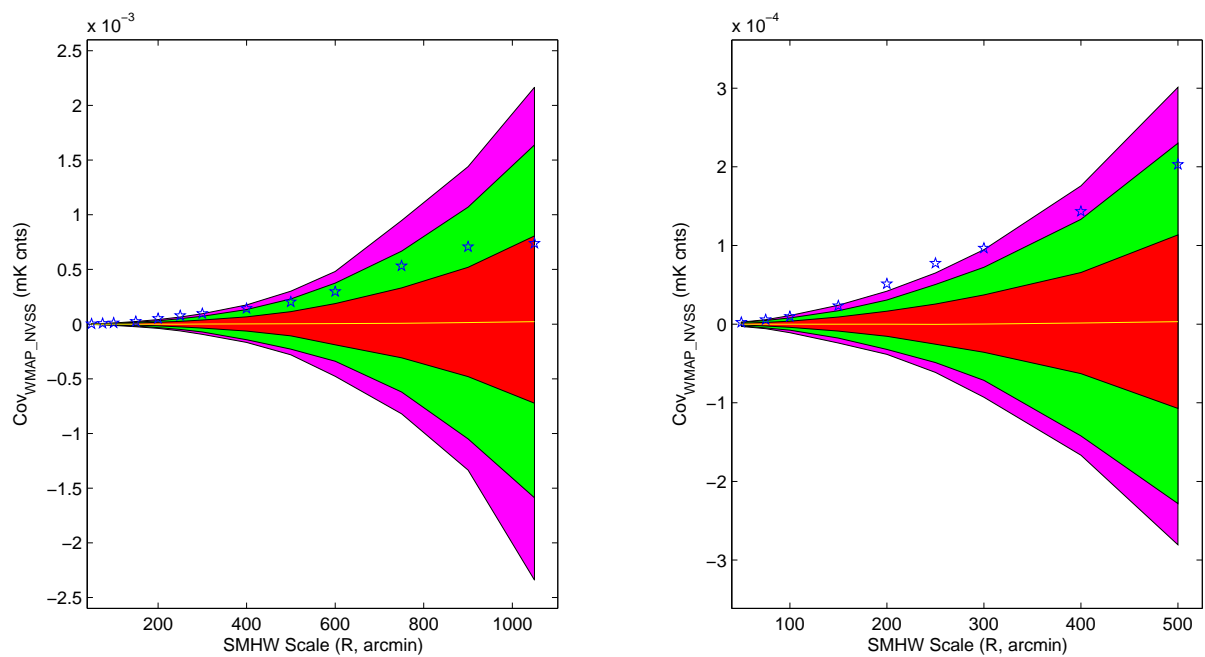}
	\caption{\label{fig.detection1}Cross-correlation in
	\emph{wavelet space} of the WMAP and NVSS data (stars). 
	Acceptance intervals for the 32\% (inner interval), 5\%
	(middle interval) and 1\% (outer interval) significance levels
	are also plotted. The mean value of the cross-correlation for
	the 1000 simulations is also given (solid line). The plot is
	magnified for the lowest scales in the right panel.}
	\end{center}
\end{figure*}

\begin{figure*}
	\begin{center}
	\includegraphics[width=8cm]{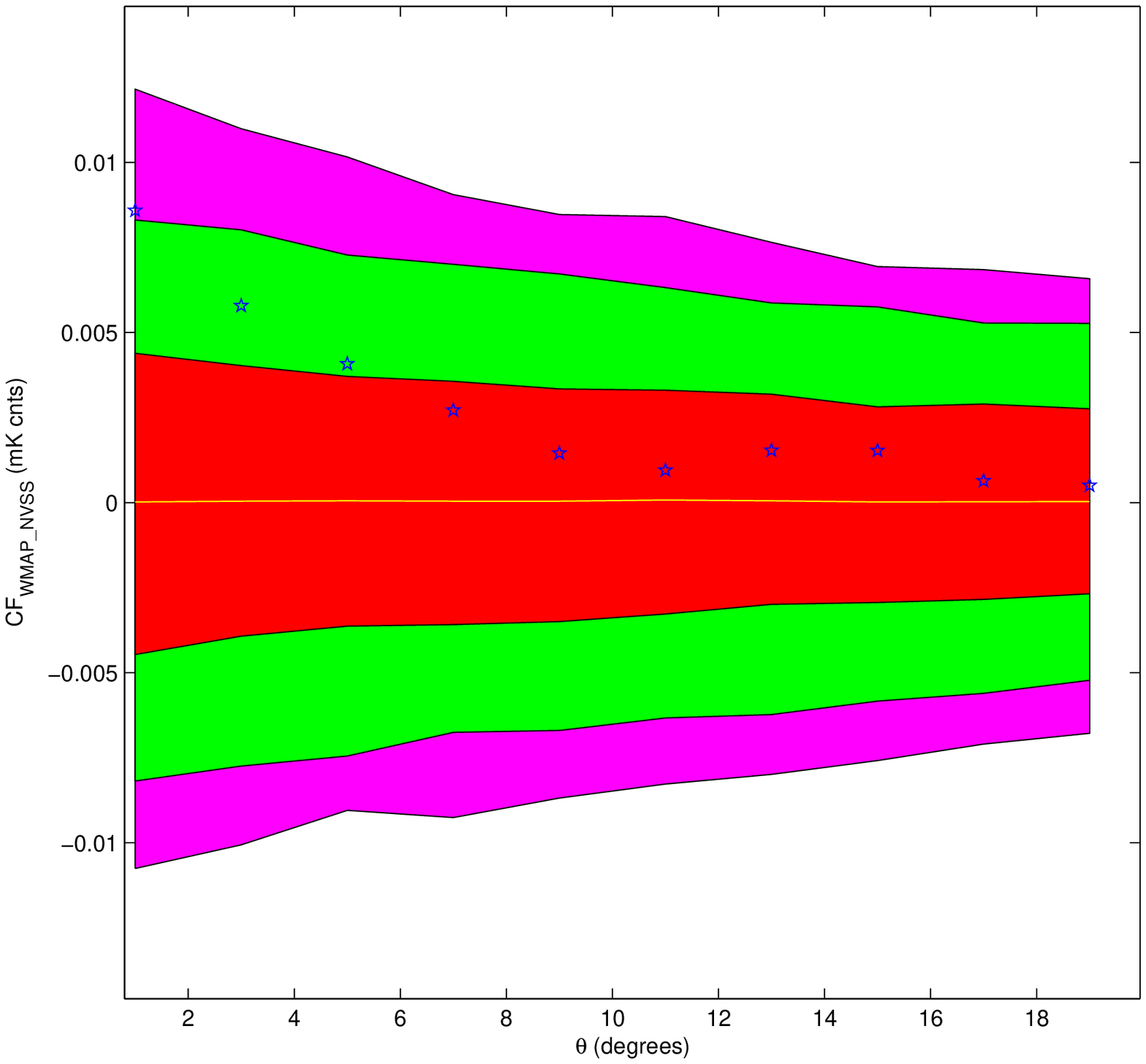}
	\includegraphics[width=8cm]{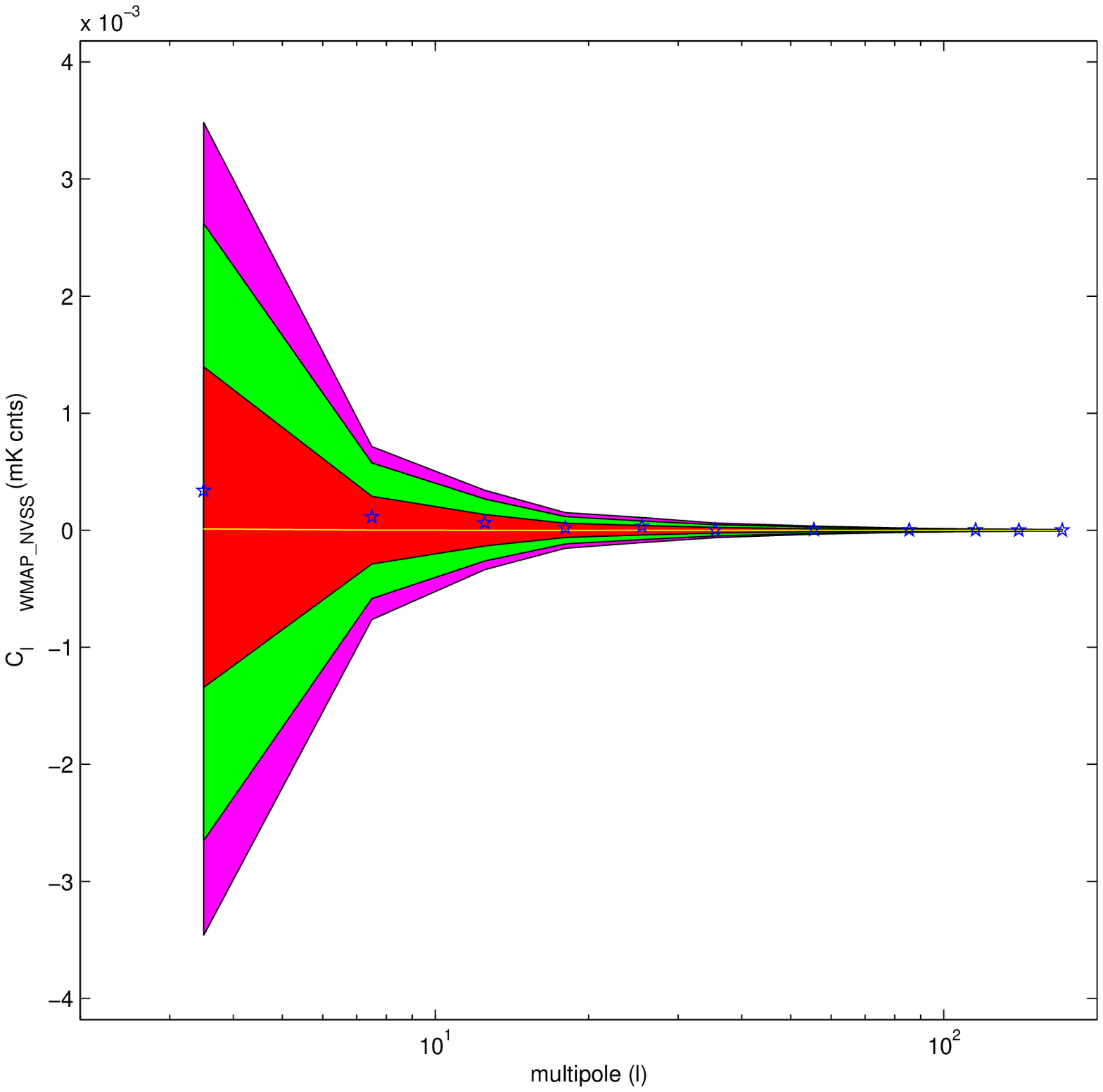}
	\caption{\label{fig.detection2}Cross-correlation in \emph{real} (left panel) and \emph{harmonic} (right panel)
\space of the WMAP and NVSS data (stars). 
Acceptance intervals for the 32\% (inner interval), 5\% (middle interval) and 1\% (outer interval) 
significance levels are also plotted. The mean value of the cross-correlation for the 1000 simulations 
is also given (solid line).}
	\end{center}
\end{figure*}

\begin{figure*}
	\begin{center}
	\includegraphics[angle=270, width=16cm]{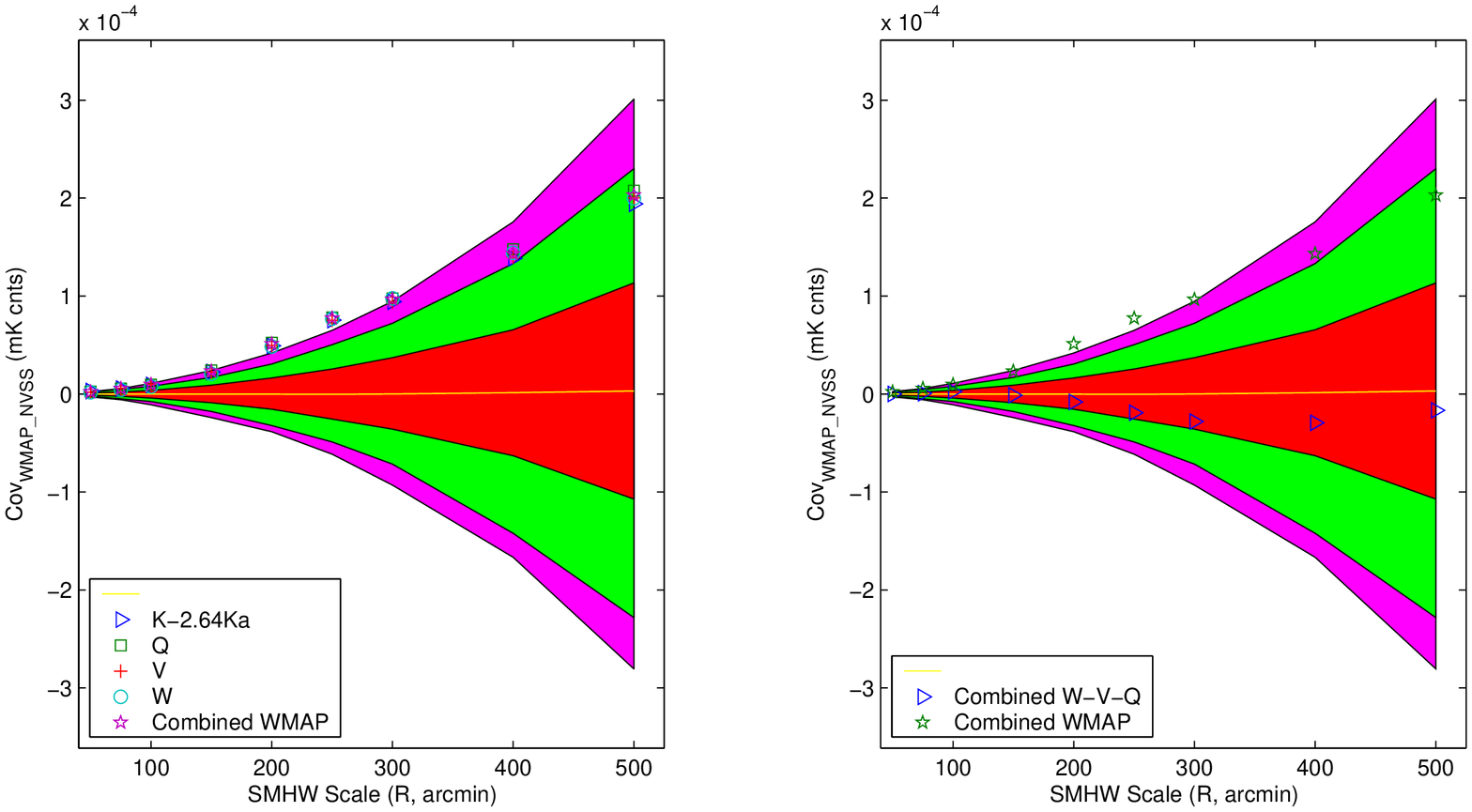}
	\caption{\label{fig.syst2} The left panel shows the CMB-nearby galaxy density correlation along the WMAP frequency range (W, V, Q, K - 2.64Ka and combined WMAP maps). The right panel shows the covariance between the noisy W-V-Q map and NVSS. The acceptance intervals are the same as the ones in Figure~\ref{fig.detection1} since they are dominated by the cosmic variance.}
	\end{center}
\end{figure*}

The results obtained from the cross-correlation of WMAP and NVSS are presented in this Section. 
In Subsection~\ref{subsec:res1} we show the ISW detection at $\approx 3.3\sigma$ provided by the CSMHW.
The curves obtained for the CCF and the CAPS are also presented.
In Subsection~\ref{subsec:res2} we compare the obtained cross-correlation with different 
theoretical curves provided by different $\Lambda$CDM cosmological models, in order to constrain the value 
of $\Omega_{\Lambda}$. The performance of the three techniques (CAPS, CCF and CSMHW) is compared, 
showing that all of them provide compatible limits at $1\sigma$ CL, being the CAPS and the CSMHW better
than the CCF in order to reject a zero dark energy value. Both of then discard values of $\Omega_\Lambda < 0.1$ with
a significance $> 3\sigma$. In particular, the CAPS detect the presence of dark energy at $3.5\sigma$, which
is the highest detection reported up to date by using the ISW.
Finally, in Subsection~\ref{subsec:res3} we extend the analysis 
in order to constrain the value of dark energy density and the equation of state parameter, finding again
compatible values at $1\sigma$ for the three statistics.

\subsection{The Detection of the ISW}
\label{subsec:res1}

We have computed the SMHW coefficients of WMAP and  NVSS maps at
different scales, from 50 arcmin to $17.5^\circ$ (which implies an
approximate size in the sky from $1.5^\circ$ to $35^\circ$)
and we have studied their cross-correlation.
Acceptance intervals for the correlated signal at certain
significance levels $\alpha$ (32\%, 5\% and 1\%) have
been established at each scale using simulations. These acceptance
intervals contain a probability  
$1 - \alpha$ and the remaining probability is the same above and below
the interval (i.e. $\alpha/2$).  
The acceptance intervals have been determined by studying the CSMHW
${Cov}_{_{{\tiny{WN}}}} (R)$ at  
each scale independently and calculating the corresponding
percentiles. 

In Figure~\ref{fig.detection1} we present the ISW detection. Scales
$R_5$, $R_6$ and $R_7$ are outside the acceptance interval at 1\%
significance level (in particular, by increasing the number of
simulations up to 10000, the scales $R_5$ and $R_6$ provide a
detection at $\approx 3.3\sigma$), representing a clear
detection of the cross-correlation of the WMAP and NVSS data.
These scales correspond to a size in the sky around $6^\circ$ -- $8^\circ$, which is in 
very good agreement with the theoretical detection predicted in Section~\ref{sec.motivation} 
and also with Afshordi (2004), that predicts an optimal scale for the ISW between 
$2^\circ$ -- $10^\circ$.

For comparison, the detection curves for the CCF (left panel) and the CAPS (right panel) are given in  
Figure~\ref{fig.detection2}. The CCF provides a detection at $\approx 95\%$, in good agreement with the
curves obtained by Boughn \& Crittenden (2004) and Nolta et al. (2004). For a better visualization,
the CAPS has been plotted after averaging multipoles in 11 intervals. As expected (see Figure~\ref{fig.s2n}), 
all the measured points are inside the acceptance interval at 32\% significance level.
We remark, as said in Section~\ref{sec.motivation}, that the CSMHW and the CCF curves are strongly correlated,
whereas the correlation along the CAPS values is very small. This indicates that, by taking into account
all the data and all the correlations in the proper way, the three techniques could provide a similar
detection of the ISW, although differences are expected due to the presence of the mask. 
This will be considered in the next Subsections, when constraints in the
nature of the dark energy are determined. 
Besides, due to the wavelet properties, almost the total signal of the ISW can
be concentrated in a very specific scale range.

We have checked that this cross-correlation signal is not caused by systematic effects. 
We focus in the CSMHW detection, since is the one with the highest significance.
In particular, the NVSS map has been also cross-correlated (independently) with each 
one of the WMAP receivers (Q1 and Q2 at 40.7 GHz, V1 and V2 at 60.8, W1, W2, W3 and W4 at 93.5 GHz). 
We found, for all the cases, the same CSMHW as for the combined WMAP map. 
We have also tested that the cross-correlations for the maps Q1-Q2, V1-V2 and W1-W2+W3-W4
with the NVSS data are compatible with simulations.
For all these 
noisy maps, the CMB and foregrounds contributions are completely removed, showing no 
cross-correlation with the NVSS data. Hence, we conclude that all the WMAP receivers are 
producing the same cross-correlation signal and that it is not due to noisy artifacts.

We have also checked that the cross-correlation signal is not due to foreground emissions. 
In particular, the expected positive correlation between the radio sources presented both
in the NVSS and the WMAP maps is considered. This correlation is expected to show a
frequency dependence, reflecting the emission law of the point source population
(e.g. Toffolatti et al. 1998). 
The cross-correlation curves for the different WMAP frequency channels 
(Q at 40.7 GHz, V at 60.8 GHz and W at 93.5 GHz) are shown in Figure~\ref{fig.syst2}. We have also considered the 
cross-correlation curve for the map K-2.64Ka, where K and Ka are the lowest frequency WMAP 
channels at 22.8 and 33.0 GHz, respectively. These channels are clearly contaminated by 
synchrotron emission. However, an additional CMB map can be generated by subtracting the Ka map 
from the K one, multiplying the first one by a factor of 2.64. This number corresponds to the 
expected increment of the synchrotron emission from 33 to 22.8 GHz\footnote{A power law is 
assumed for the frequency dependence of the synchrotron emission: 
$T_{syn}(\nu) \propto T_{syn}(\nu_0)(\nu/\nu_0)^{-2.7}$, as proposed by Bennett et al. (2003b).}. 
As seen in the left panel of Figure~\ref{fig.syst2}, the same cross-correlation curve is 
obtained from the whole WMAP frequency range, showing not frequency dependence at all. 
Moreover, the cross-correlation curve for the W-V-Q map (free of CMB signal but with a 
clear foreground contribution) is perfectly compatible with zero (Figure~\ref{fig.syst2}, right panel). 
Hence, we conclude that the cross-correlation is only due to the CMB, with a negligible 
contribution from foregrounds, as expected for the ISW signal.

\subsection{Constraints in $\Omega_{\Lambda}$}
\label{subsec:res2}
\begin{figure*}
	\begin{center}
	\includegraphics[width=8cm]{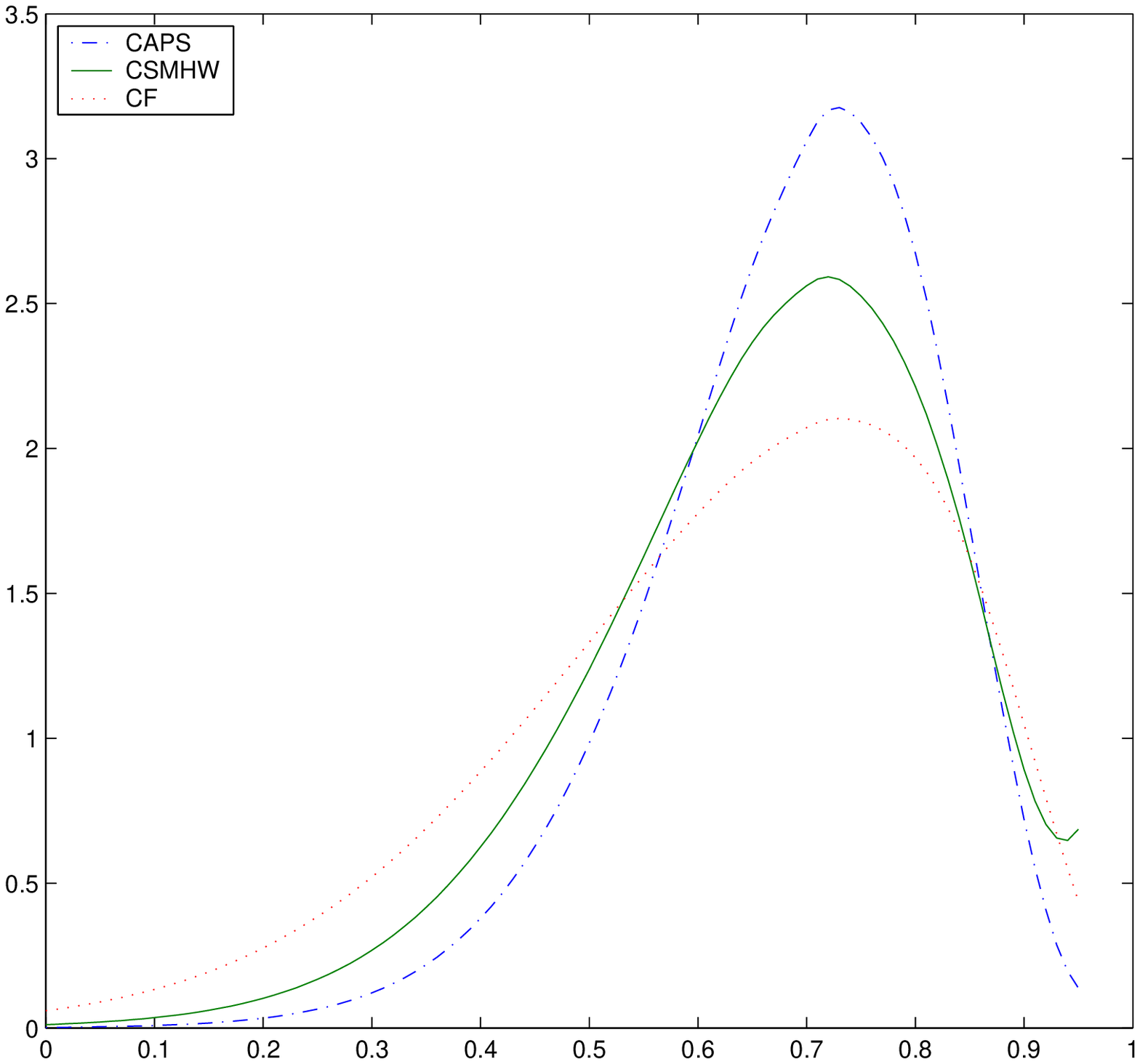}
	\includegraphics[width=8cm]{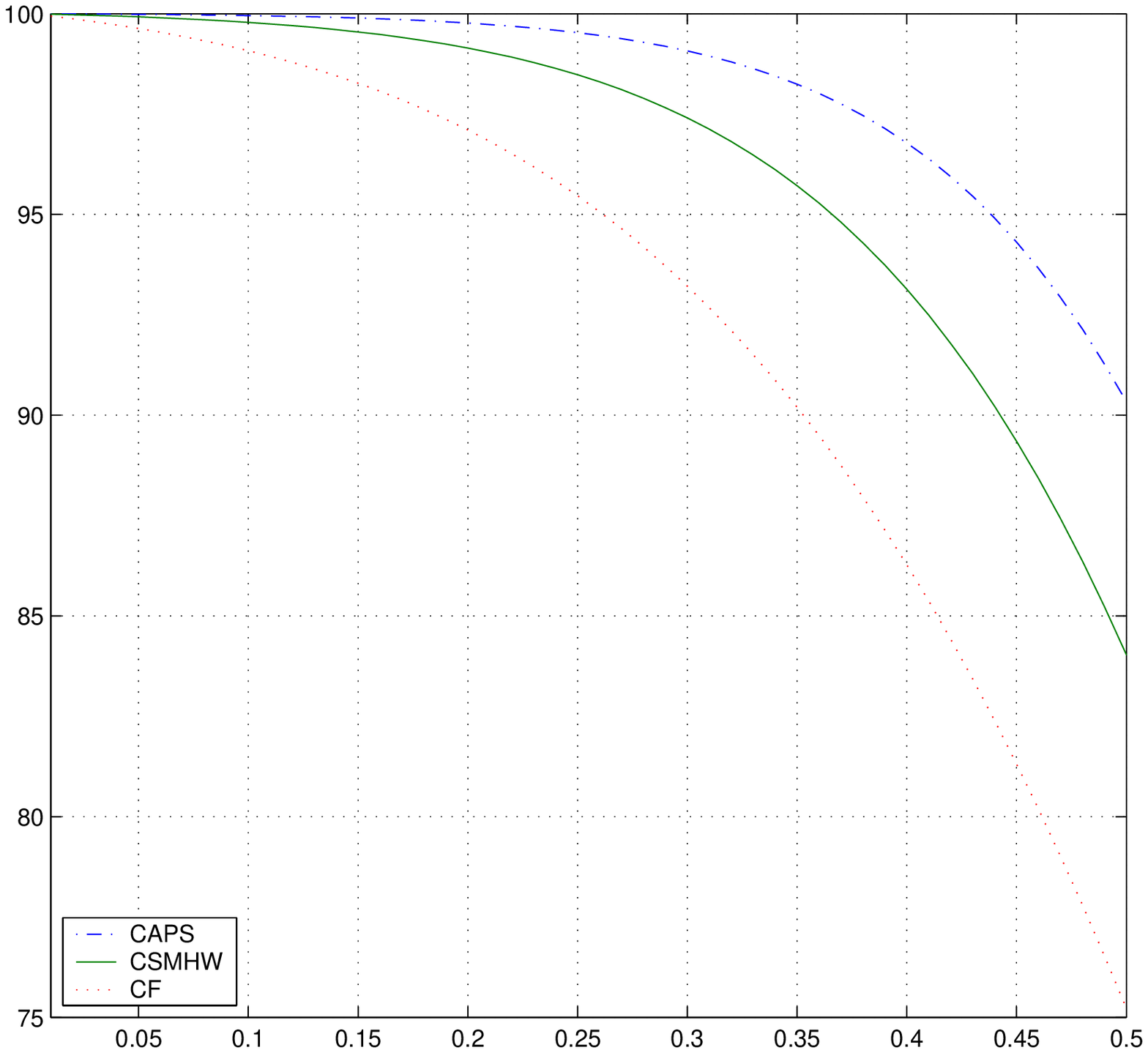}
	\caption{\label{fig.lambda.like}In the left panel, the
         likelihoods of the considered statistics (CAPS, CCF and CSMHW)
         are plotted as a function of $\Omega_\Lambda$. In the
         right panel we show the cumulative probability of 
         having an $\Omega_{\Lambda}$ greater than a certain value.}

	\end{center}
\end{figure*}

By comparing the measured detection with the theoretical curves for different cosmological models, 
limits on the cosmological parameters can be established.
In this first approach, we are interested in putting constraints on the amount 
of vacuum energy density $\Omega_\Lambda$, assuming a standard inflationary universe
with $w = -1$, without 
redshift evolution. We have explored $0 < \Omega_\Lambda < 0.95$. 
For the bias parameter, we have considered
the analysis of the NVSS catalogue (using the auto-correlation function) 
made by Boughn \& Crittenden (2002, 2004) adopting the Dunlop \& Peacock (1990) RLF1 model. 
They found that observations favor bias values of $b$ = 1.3 -- 1.6. 
For our analysis (where $H_o = 0.72$ and $\Omega_m = 0.29$),
$b = 1.6$ is the best value. 
In any case, we have checked that 
variations of the bias in the interval $1.4 \le b \le 1.8$ do not change the results. 
The Dunlop \& Peacock (1990) RLF1 model was also used to describe the $dN/dz$ distribution
since it fits well the NVSS galaxy auto-correlation, as it was already shown by 
Boughn \& Crittenden (2002, 2004) and Nolta et al. (2004).

In practice, the observational curves for the CAPS, CCF and CSMWH
(given by Equations~\ref{eq.obcls1},~\ref{eq.obscross-cf}
and~\ref{eq.estimator}) are compared
with the theoretical predictions
(Equation~\ref{eq.obcls2},~\ref{eq.cross-cf} and~\ref{eq.theo}). 
For each statistic, a generalized $\chi^2$ is defined:
\begin{equation}
\label{eq.chi2.1}
\chi^2(p| b) =
\Delta_{cross}^t(p | b; a_i)
{C(a_i, a_j)}^{-1}
\Delta_{cross}(p | b; a_j)
\end{equation}
where the parameter $p \equiv \Omega_\Lambda$ and
$\Delta_{cross}(a) = \left( {{cross}^{theo}}_{_{{\tiny{WN}}}} (p | b; a) - {cross}_{_{{\tiny{WN}}}} (a) \right)$ 
is the difference between the observed CAPS/CCF/CSMHW and the theoretical prediction 
at scale $a$ (i.e., the CSMHW $R$ scale or the angular scale $\theta$ of 
the CCF or the multipole $\ell$ of the CAPS).
The matrix $C(a_i, a_j)$ is the correlation matrix of the WMAP and NVSS CAPS/CCF/CSMHW 
and it is obtained from 1000 simulations. Assuming that each $C(a_i, a_j)$ is Gaussian distributed, a likelihood
$L \propto exp(-\chi^2 / 2)$ can be calculated. The likelihood obtained for each statistic is represented in
Figure~\ref{fig.lambda.like} (left panel). At $1\sigma CL$, we obtain compatible constraints for the
vacuum energy density: $\Omega_{\Lambda} = 0.72^{+0.12}_{-0.15}$ 
(for CSMHW), $\Omega_{\Lambda} = 0.73^{+0.12}_{-0.18}$ (for CCF)\footnote{Notice that
the likelihood obtained for the CCF is slightly different from the one given in Nolta et al. (2004, Figure~4d),
although both provide consistent limits at $1\sigma$ CL.
There are differences that can explain it. First, in Nolta et al. (2004) the theoretical
ISW effect is removed from the simulations to estimate the covariance matrix.
Second, the estimate of the CCF is different in Nolta et al. (2004), since
each pixel is weighted by taking into account the number of unmasked 
pixels at the original resolution.} and 
$\Omega_{\Lambda} = 0.73^{+0.11}_{-0.14}$ (for CAPS).

On the right panel of Figure~\ref{fig.lambda.like} we represent (for
each statistic) the cumulative
probability of having an  $\Omega_{\Lambda}$ 
greater than a certain value. This quantity provides us an
alternative method to estimate the significance of the ISW detection.
These plots shows how the CAPS and the CSMHW 
are more powerful than the CCF for this purpose.
There are several reasons for understanding this behaviour.
On the one hand, the CAPS statistic has the important advantage that,
apart from the mask effect, there are not correlations between multipoles, which
is always a desirable property to build the chi-square, although all the available multipoles
contribute in a similar way. On the other hand, although the CSMHW statistic has correlations
among different scales,  most of the signal is concentrated in a narrow angular range 
(that, in this case, is around several degrees), i.e. the
most favourable condition to use wavelets. On the contrary, the CCF one
has correlations among angular scales
and, in addition, its maximum significance is smaller than the SMHW one.
In particular, the CAPS provides $\Omega_{\Lambda} > 0.1$ at a significance of $\approx 3.5\sigma$,
which is the highest detection of the vacuum dark energy (based on the ISW) reported up to date.

The reason of those differences in the significance of the detection is due to the presence of a mask.
In the case of full-sky data sets, a maximum likelihood method applied to the three quantities should give very similar results, since
each of them can be expressed as a linear combination of any of the other two.

\begin{figure*}
	\begin{center}
	\includegraphics[width=5.3cm]{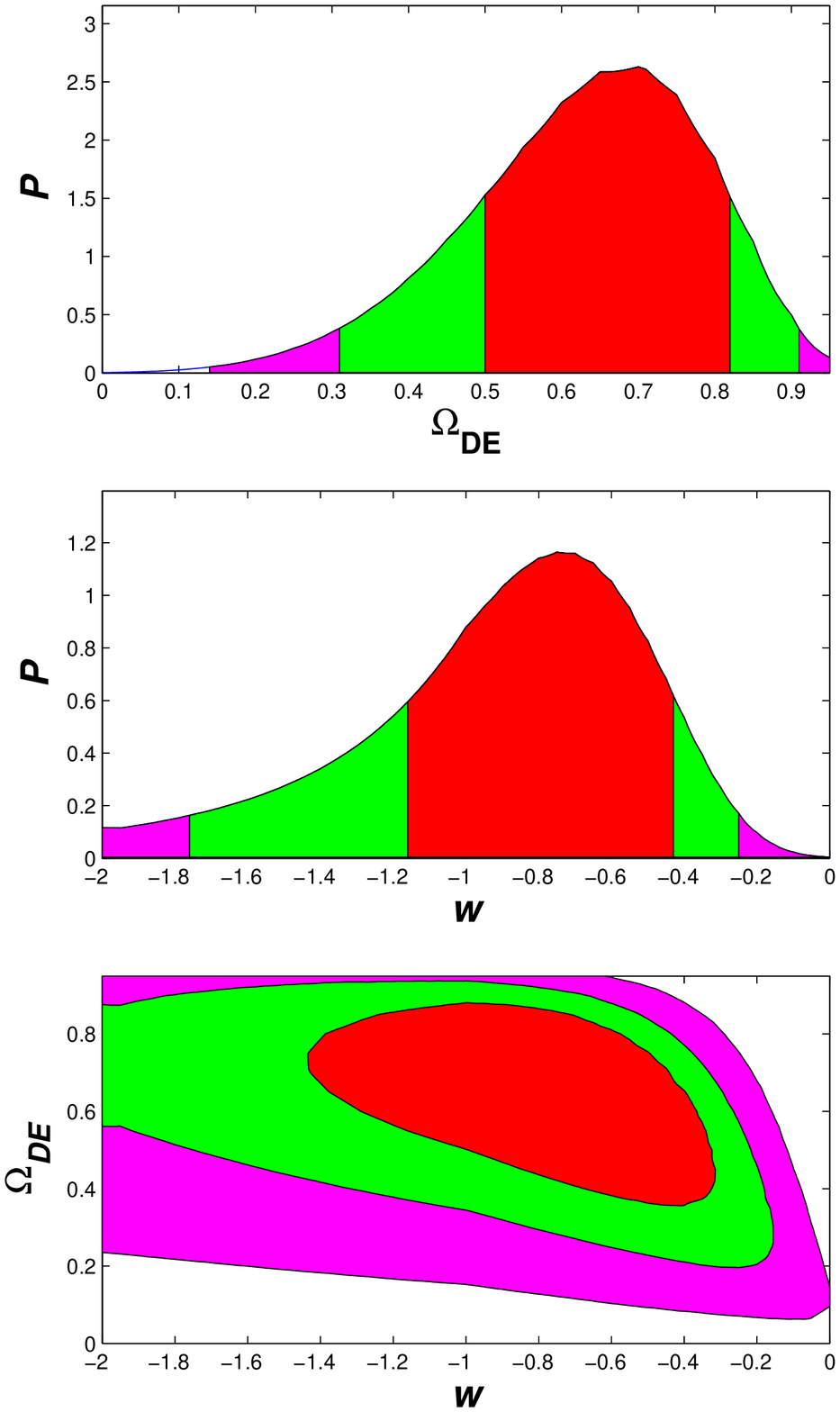}
	\includegraphics[width=5.3cm]{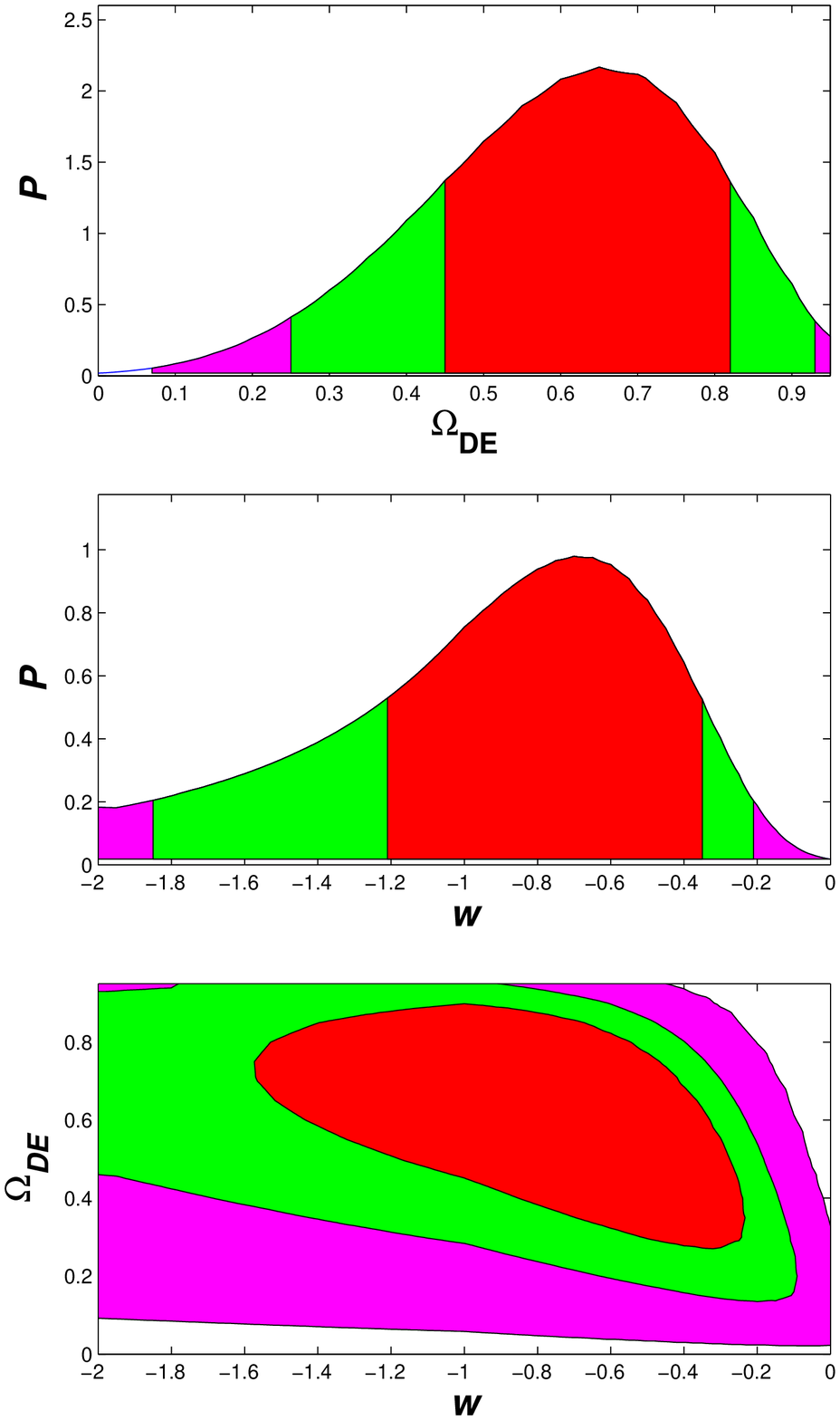}
	\includegraphics[width=5.3cm]{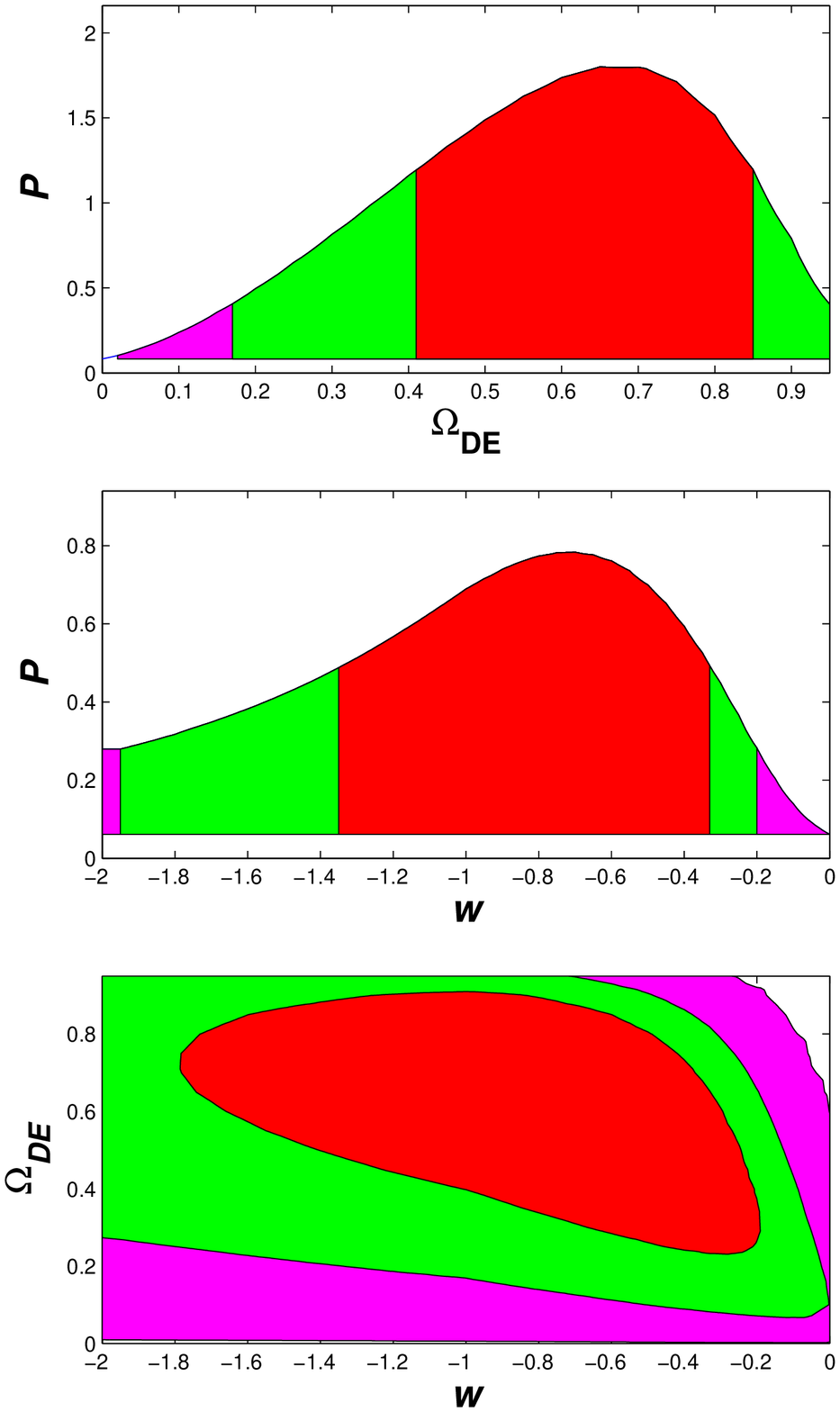}
	\caption{\label{fig.prob} The logarithm of the marginalized pdf for $\Omega_{DE}$ (upper) and $w$ (middle) are presented. The 2D likelihood is also provided (bottom). The errors in the parameter estimation at $1\sigma$, $2\sigma$ and $3\sigma$ are also plotted. First column shows the limits given by the CAPS, the second ones is for the CSMHW and the last one is for the CCF.}

	\end{center}
\end{figure*}

\subsection{Constraints in $\Omega_{DE}$ and $w$}
\label{subsec:res3}

In the present section, we are interested in putting constraints in the nature of the dark energy.
In particular, the amount of the dark energy density $\Omega_{DE}$ 
(and so, the matter density, since we impose a flat universe) 
and the equation of state parameter $w$ are studied.
As it was done in the previous Section, 
we put limits in these cosmological parameters by comparing the
measured WMAP and NVSS cross-correlation with the theoretical curves
obtained from different cosmological models. The explored range for the dark energy density is 
$0 < \Omega_{DE} < 0.95$, whereas the one for the equation of state
parameter is $-2.0 < w < 0$.

In this situation, a generalized $\chi^2$ can be defined as in Equation~\ref{eq.chi2.1},
with parameters $p \equiv (\Omega_{DE}, w)$.
As it was said in the Subsection~\ref{subsec:res2}, the elements of the correlation matrix 
are assumed to be Gaussian, hence we can define a likelihood as $L \propto exp(-\chi^2 / 2)$. A joint 
pdf of the parameters $f(\Omega_{DE}, w)$ can be calculated by normalizing the likelihood 
in the whole parameter space. The 2D likelihood is shown in Figure~\ref{fig.prob} (bottom panel).
It is straightforward to calculate the marginalized pdf distributions $f(\Omega_{DE})$ and $f(w)$. 
These marginalized pdfs are plotted in Figure~\ref{fig.prob} in the upper ($f(\Omega_{DE})$) 
and in the middle ($f(w)$) panels. The results for the CAPS, CSMHW and CCF are given in columns 
1, 2 and 3, respectively.

We obtain at $1\sigma$ CL: $\Omega_{DE} = 0.70^{+0.12}_{-0.20}$ and $w
= -0.75^{+0.32}_{-0.41}$ with the CAPS;
$\Omega_{DE} = 0.70^{+0.15}_{-0.29}$ and $w = -0.75^{+0.42}_{-0.60}$
with the CCF; and  $\Omega_{DE} = 0.65^{+0.17}_{-0.20}$ and $w =
-0.70^{+0.35}_{-0.50}$ with the CSMHW.

The limits in the dark energy density are compatible with the vacuum
energy density calculated in the previous
section. The constraints in $w$ are less restrictive.
The standard inflationary concordance model with $\Omega_{DE} \approx
0.71$ and $w = -1$ is perfectly compatible within the $1\sigma$ CL.
As explained in the previous Subsection, the differences in the results 
provided by the three statistics arise because of the presence of the mask.

\section{Conclusions}
\label{sec.conclusion}

We have performed the cross-correlation of the WMAP first-year data with the radio 
galaxy distribution traced by the NVSS catalogue, in the \emph{harmonic}, \emph{real} and \emph{wavelet} \emph{spaces}.
This is the first application of wavelet techniques to study the cross-correlations 
between the CMB and the nearby universe. We found a clear cross-correlation at scales 
in the sky around $\theta \approx 7^\circ$ with a significance $\approx  3.3\sigma$, in good agreement
with theoretical predictions. Wavelets provide a better detection level respect to 
statistics based on spatial and harmonics space. This is due to the important advantage of
wavelets to concentrate, in a particular and narrow scale range, almost the total signal of 
the ISW. On the other hand, the CAPS and the CCF provide a similar detection level only if
the whole information spread along all the multipoles/angles is properly combined and no mask is present.

Even more, by using the CSMHW, it is not required the calculation of different theoretical models, 
which are necessary to properly combine the whole information in the CAPS and CCF cases.
The direct detection of the ISW obtained with the CSMHW is independent of the particular values 
for different parameters, as the bias, the point source redshift distribution, the equation of the estate parameter
of the dark energy, etc. The ISW detection at $\approx  3.3\sigma$ is the highest significance level of the ISW 
reported up to date.
By calculating the CSMHW for different combinations of the WMAP receivers, we have proven that this cross-correlation 
signal is not caused neither by systematic effects nor foreground contamination.

However, the comparison of the measured signal with the expected theoretical cross-correlation curves
for different cosmological models, is needed to put constraints on the cosmological parameters.
In particular, we have put limits to the amount of vacuum energy density $\Omega_\Lambda$ (assuming $w \equiv 1$), 
obtaining compatible constraints at $1\sigma$ for the three different
techniques: at $1\sigma CL$, $\Omega_{\Lambda} = 0.72^{+0.12}_{-0.15}$ 
(for CSMHW), $\Omega_{\Lambda} = 0.73^{+0.12}_{-0.18}$ (for CCF) and 
$\Omega_{\Lambda} = 0.73^{+0.11}_{-0.14}$ (for CAPS).
Even more, the CAPS provides $\Omega_{\Lambda} > 0.1$ at a significance of $\approx 3.5\sigma$,
which is the highest detection of the vacuum dark energy (based on the ISW) reported up to date.
We have used the bias parameter $b = 1.6$ already calculated by Boughn \& Crittenden (2002) for the 
NVSS data and the Dunlop \& Peacock (1990) RLF1 model for the $dN/dz$ distribution. Our results do 
not change for bias values in the interval  $1.4 \le b \le 1.8$.

We have also tested the ability of the cross-correlation to put constraints on the nature on the
dark energy by studying the amount of the dark energy $\Omega_{DE}$ and the equation of state parameter $w$.
We obtain  $\Omega_{DE} = 0.70^{+0.12}_{-0.20}$ and $w = -0.75^{+0.32}_{-0.41}$ (at $1\sigma$ CL)
with the CAPS, $\Omega_{DE} = 0.70^{+0.15}_{-0.29}$ and $w = -0.75^{+0.42}_{-0.60}$ (at $1\sigma$ CL)
with the CCF and  $\Omega_{DE} = 0.65^{+0.17}_{-0.20}$ and $w = -0.70^{+0.35}_{-0.50}$ (at $1\sigma$ CL)
with the CSMHW.
Our estimation of the equation of state of the dark energy is the first one made through the 
cross-correlation of the CMB and the nearby galaxy density distribution. It provides an independent 
determination from that made by the WMAP team (Spergel et al. 2003) using CMB and LSS and by 
other groups (Caldwell \& Doran 2004, Melchiorri 2004 and Corasaniti et al. 2004 and references 
therein) using, in addition to CMB and LSS, data coming from SN-Ia and/or BBN.

We remark that the differences found in the determination of the cosmological parameters 
are due to the presence of a mask. In the case of full-sky data sets, a maximum likelihood method applied 
to the three quantities should give very similar results, since each of them can be expressed as a linear 
combination of any of the other two.

Finally, in this paper we have shown that wavelets are a very promising tool for studying the 
correlation of the CMB with LSS data. Other effects like Sunyaev-Zeldovich or the contribution 
from point sources could also be studied having an appropriate LSS tracer.

\section*{Acknowledgments}
We thank Martin Kunz for useful discussions on dark energy properties.
We also acknowledge Jos\'e L. Sanz and Bel\'en Barreiro for very interesting comments.
We are very thankful to Fernando Atrio-Barandela and Carlos Hern\'andez-Monteagudo,
for very useful comments regarding the NVSS data. We acknowledge 
the financial support provided through the European
Community's Human Potential Program under contract
HPRN-CT-2000-00124, CMBNET, and
partial financial support from the Spanish MEC projects 
ESP2002-04141-C03-01 and ESP2004-07067-C03-01.
PV thanks IN2P3 (CNRS) for a post-doctoral contract.
We acknowledge the use of LAMBDA, support for which is provided by the NASA Office of Space Science.
This work has used the software package HEALPix (http://healpix.jpl.nasa.gov) developed by
K. M. G\'orski, E. F. Hivon, B. D. Wandelt, J. Banday, F. K. Hansen and M. Barthelmann. 
We acknowledge the use of the software package CMBFAST (http://www.cmbfast.org) 
developed by U. Seljak and M. Zaldarriaga.

\bsp

\end{document}